\documentclass[twocolumn,twocolappendix]{aastex631}  
\usepackage{verbatim}
\usepackage{graphicx}
\usepackage{xcolor}
\usepackage{booktabs}
\usepackage{xspace}
\usepackage{placeins}
\usepackage{longtable}
\usepackage{threeparttablex}
\usepackage{natbib}
\usepackage{txfonts}\usepackage{twoopt}
\bibpunct{(}{)}{;}{a}{}{,}
\usepackage{enumitem}
\defcitealias{bronner2025}{Paper~I}

\newcommand{\msun}{{\rm M}_{\odot}}
\newcommand{\lsun}{{\rm L}_{\odot}}
\newcommand{\rsun}{{\rm R}_{\odot}}

\newcommand{\eg}{e.g.\@\xspace}

\shorttitle{Exploding pulsating red supergiants: application to SN~2023ixf and SN~2024ggi}
\shortauthors{E. Laplace, V.~A.~Bronner et al.}

\graphicspath{{./}{figures/}}

\begin{document}

\title{Pulsations change the structures of massive stars before explosion: interpreting SN~2023ixf and SN~2024ggi}

\correspondingauthor{Eva Laplace and Vincent Bronner}
\email{eva.laplace@kuleuven.be, vincent.bronner@h-its.org}

\author[0000-0003-1009-5691]{Eva~Laplace}
\affiliation{
        Institute of Astronomy, KU Leuven, Celestijnenlaan 200D, B-3001 Leuven, Belgium, 
}
\affiliation{
Leuven Gravity Institute, KU Leuven, Celestijnenlaan 200D, box 2415, 3001 Leuven, Belgium
}
\affiliation{
        Anton Pannekoek Institute of Astronomy, University of Amsterdam, Science Park 904, 1098 XH Amsterdam, The Netherlands}
\affiliation{Heidelberger Institut f{\"u}r Theoretische Studien, Schloss-Wolfsbrunnenweg 35, 69118 Heidelberg, Germany}

\author[0000-0002-7624-2933]{Vincent~A.~Bronner}
\affiliation{Heidelberger Institut f{\"u}r Theoretische Studien, Schloss-Wolfsbrunnenweg 35, 69118 Heidelberg, Germany}
\affiliation{Universit\"{a}t Heidelberg, Department of Physics and Astronomy, Im Neuenheimer Feld 226, 69120 Heidelberg, Germany}

\author[0000-0002-5965-1022]{Fabian R.~N.~Schneider}
\affiliation{Heidelberger Institut f{\"u}r Theoretische Studien, Schloss-Wolfsbrunnenweg 35, 69118 Heidelberg, Germany}
\affiliation{Astronomisches Rechen-Institut, Zentrum f{\"u}r Astronomie der Universit{\"a}t Heidelberg, M{\"o}nchhofstr.\ 12-14, 69120 Heidelberg, Germany}

\author[0000-0002-8338-9677]{Philipp~Podsiadlowski}
\affiliation{London Centre for Stellar Astrophysics, Vauxhall, London}
\affiliation{University of Oxford, St Edmund Hall, Oxford, OX1 4AR, United Kingdom}
\affiliation{Heidelberger Institut f{\"u}r Theoretische Studien, Schloss-Wolfsbrunnenweg 35, 69118 Heidelberg, Germany}

\accepted{26 January 2026}
 
\begin{abstract} 
    Massive red supergiants (RSGs) are known to become hydrodynamically unstable before they explode. Still, the vast majority of supernova (SN) models assume RSG progenitors in hydrostatic equilibrium. Here, we follow the hydrodynamic evolution of RSGs with different masses and the development of radial envelope pulsations. Pulsations significantly alter the observable pre- and post-SN properties, and their importance increases substantially as a function of initial mass. We demonstrate that inferring core masses, let alone initial masses, from a single pre-SN luminosity and effective temperature of high-mass RSGs is inadvisable, as these can vary by an order of magnitude during the pulsation. We find that pulsations can naturally lead to ``early-excess" emission in SN light curves and to variations in early photospheric velocities, which can help break degeneracies in type-II SNe. We compare to SN~2023ixf and SN~2024ggi, for which pulsating RSG progenitors were reported. We demonstrate that the pre- and post-SN characteristics of SN~2023ixf agree very well with our exploding pulsating RSG model and exhibit meaningful differences from hydrostatic models. The data coverage is insufficient to break all degeneracies. We find insufficient evidence for the claimed pulsation period of the SN~2024ggi progenitor, as it matches Spitzer’s orbital period. This study underscores the importance of hydrodynamical pre-SN stellar models, in particular for massive stars from $\gtrsim 15\,\msun$. It implies an important shift in our understanding of the last stages of massive star evolution, the interpretation of pre-SN properties, the connection between SNe and their progenitors, and the missing RSG problem.
\end{abstract}


\section{Introduction}
\label{sec:intro}
The final life of massive stars is poorly constrained. This last phase of the evolution after core helium burning lasts only on the order of thousands to tens of thousands of years, which means that it is rare to find stars in this evolutionary stage observationally. Their properties in this last stage are particularly uncertain. Single stars with masses of $\approx8\,\msun$ and up to $\approx25\,\msun$ are expected to end their lives as luminous red supergiants (RSGs). While supernova (SN) observations point to growing evidence for dense circumstellar material (CSM) near SNe inferred from narrow, flash-ionization emission features in their early spectra \citep{gal-yam_2014_Wolf-Rayet, bruch_2021_large}, early excess in SN light-curves (LCs) \citep{morozova_2018,hinds_2025_inferring}, and late-time radio, infrared, and X-ray emission of nearby SNe \citep{chugai_optical_2007,fransson_2014_high}, observationally-inferred mass loss rates of RSGs are orders of magnitude lower \citep{davies2018a,beasor2020a,yang2018a,antoniadis2024a,yang_2023_evolved,decin_2024_ALMA}. In addition, massive RSGs are highly variable \citep{kiss2006a, percy2014a}, likely due to radial envelope pulsations \citep{stothers1969a,stothers1971a, heger1997a,yoon2010a}. Their period-luminosity (PL) relations have been used to infer the distances to other galaxies \citep[\eg][]{jurcevic2000a}. However, most supernova explosion models rely on progenitors assumed to be in hydrostatic equilibrium.

The discovery of SN~2023ixf \citep{Itagaki_2023_TNS} and its pulsating red supergiant progenitor \citep[\eg][]{kilpatrick2023a} put the validity of this assumption into question. This hydrogen-rich (Type II) SN \citep{perley_2023_classification}, located in the Pinwheel galaxy, M~101, at a distance of $6.85\pm0.15\,\rm{Mpc}$ , is among the closest SNe to have occurred in the past decade. At the location of the supernova, archival searches revealed a variable red supergiant progenitor with a pulsation period of $1100$ d \citep{kilpatrick2023a, vandyk2024a,jencson2023a,soraisam2023a,niu2023a,qin2024a,ransome2024a,xiang2024a}. Given the close proximity of the supernova, it has led to a wealth of observations across the electromagnetic spectrum. Several studies reported evidence for CSM close to the SN \citep[\eg][]{jacobson-galan_2023_core-collapse,jencson2023a,smith2023,zimmerman2024a}. In addition, there are several signs of possible large-scale asymmetries in the ejecta, which have been interpreted as a signature of past outflows or binary interactions \citep{vasylyev_2025_spectropolarimetric,shrestha_2025_spectropolarimetry,singh2024}. For interpreting the properties of this object, few studies have taken the pulsation of the progenitor into account. A notable exception is the work by \citet{hsu_2024_one}, who proposed to only select stellar models consistent with the pulsation period to break degeneracies in the explosion properties. 

More recently, another close-by H-rich SN, SN~2024ggi was discovered \citep{chen2025a}, for which recent studies find a variable RSG progenitor \citep{xiang2024b}. These objects provide a rare glimpse into the final properties of RSGs before their explosions.

Historically, few studies have attempted to follow the hydrodynamical evolution of stars before explosion \citep{heger1997a, yoon2010a}. \citet{clayton2017a} studied the hydrodynamical properties of giants with unstable envelopes self-consistently in the context of common-envelope evolution, and \citet{clayton2018a} explored the properties of RSGs and their mass loss shortly before explosion. \citet{goldberg2020a} calculated the pulsations of a large range of massive stars. By introducing a velocity perturbation in the envelope, they triggered radial pulsation in stellar models of RSGs at the end of core carbon burning. \citet{goldberg2020a} revealed that such pulsations can affect the final structure of stars, their stellar LCs, and also their SN LCs. Following up on the work of \citet{clayton2017a,clayton2018a}, we recently computed the hydrodynamic evolution of a $15\,\msun$ RSG self-consistently \citep{bronner2025}, hereafter \citetalias{bronner2025}, and found that it naturally undergoes pulsations through a $\kappa\gamma$-mechanism \citep{heger1997a,yoon2010a,clayton2017a,suzuki2025a}. We discussed the origin and mechanism behind the pulsations and showed that they strongly affect the pre-SN structure of the RSG, leading to SNe of varying SN LC decline rates that may be classified as type II-P, short-plateau II-P, or type II-L. In this \textit{Letter}, we extend our study to multiple initial masses and show that large-amplitude pulsations can significantly change the pre-SN structure and explosion properties of RSGs. By comparing our models of pulsating RSGs to SNe with observed pulsating progenitors, we show that the observational characteristics both of the SN and pre-SN stellar LC can be understood as the consequence of naturally occurring pulsations of RSGs. We demonstrate that hydrostatic models are insufficient to describe the structure of such stars and their explosions, and emphasize the need for hydrodynamical models for understanding the final evolution and explosion properties of massive stars.

Our methods for the progenitor and explosion modeling are described in Sect.~\ref{sec:method:prog} and Sect.~\ref{sec:method:sn}, respectively. Following the same approach as in \citetalias{bronner2025}, we investigate the effect of pulsations on the final properties of massive single stars with different initial masses in Sect.~\ref{sec:results:rsg_models}. We explore the resulting SN LCs and photospheric velocities at different pulsation phases in Sect.~\ref{sec:results:sn_models}. In Sect.~\ref{sec:results:prog_23ixf_24ggi}, we check the claimed variability of the progenitors of SN~2024ggi and SN~2023ixf and demonstrate that there is no conclusive evidence for pulsations in SN~2024ggi. We then compare our models to the progenitor of SN~2023ixf in Sect.~\ref{sec:results:prog_23ixf} and to its explosion properties in Sect.~\ref{sec:results:sn_23ixf}. We discuss our findings in Sect.~\ref{sec:discussion} and present our conclusions in Sect.~\ref{sec:conclusion}.

\section{Methods}
\label{sec:method}
\subsection{Modeling red supergiant supernova progenitors}
\label{sec:method:prog}
We use the stellar evolution code \texttt{MESA}, v.10398 \citep{paxton2011a,paxton2013a,paxton2015a,paxton2018a,paxton2019a} to simulate the evolution of massive single, non-rotating stars with initial masses of $10.5$, $12.5$, and $15.0\,\msun$ at solar metallicity ($Z=0.0142$) until the onset of core-collapse. Following the method described in \citetalias{bronner2025}, we compute the hydrodynamic evolution of these stars at the end of core carbon burning and follow the formation of stellar pulsations until they reach a steady and periodic equilibrium state. In particular, we use the implicit hydrodynamic solver which makes use of artificial viscosity (for a discussion about the solver and implications of damping, see Appendix A of \citetalias{bronner2025}). The $15\,\msun$ model is described in detail in \citetalias{bronner2025}. For the $10.5$ and the $12.5\,\msun$ model we follow the same approach as in \citetalias{bronner2025} and remove the core at $3.6$ and $2.8\,\msun$ respectively, corresponding to $90\,\%$ for the helium core mass. An important factor for the pulsations is our choice for the efficiency of convection, which is computed according to mixing-length theory with an efficiency parameter $\alpha_{\rm{MLT}} = 1.8$. We simulate the optical and infrared stellar LCs of these pulsating RSGs using MARCS models \citep{gustafsson2008a} to describe the RSG spectrum and include a dust-shell model with the DUSTY code \citep{ivezic1997a,ivezic1999a} for predicting the optical and infrared stellar LCs (for a detailed description, see Section~2.3 of \citetalias{bronner2025}).\\  

\subsection{Modeling supernova light curves}
\label{sec:method:sn}
By the end of the evolution of massive stars, strong neutrino losses decouple the rapidly evolving core from the envelope \citep[e.g.,][]{woosley2002}. Core collapse may thus occur at any point during envelope pulsations. We compute the explosion of RSGs at different phases of the pulsation with the Lagrangian, one-dimensional hydrodynamical stellar explosion code \texttt{SNEC}, with the same assumptions as in \citetalias{bronner2025}. For our base assumptions of the explosion engine properties, we apply the semi-analytical neutrino-driven supernova code of \citet{muller2016a} with the same calibrations as in \citet{schneider2021a,temaj2024a,laplace_2025_written}, which results in a neutron star mass $M_{\rm{NS}}$, explosion energy $E$ (in units of Bethe, $1\,\rm{B} = 10^{51}\rm{erg}$), and mass of synthesized $^{56}\textrm{Ni}$, $M_{\rm{Ni}}$. For our default models, we assume that nickel is uniformly mixed throughout the stellar envelope until a boundary mass $M_{\rm{b}}=0.9\,M$, where $M$ is the final mass of the star. The explosion properties are summarized in Table~\ref{tab:summary-exp}. Using publicly available data, we compare our models to SN~2024ggi and SN~2023ixf.
To fit the SN data of SN~2023ixf, we construct 2016 \texttt{SNEC} models in which we vary the explosion properties ($E$ from 0.9 to 1.45B, $M_{\rm{Ni}}$ from $0.05$ to $0.06\,\msun$, $M_{\rm{b}}$ from $0.5$ to $0.99\, M$) for progenitors at different pulsation phases $\phi_{\rm{exp}}$ at the moment of explosion.

\section{Results}
\label{sec:results}
\subsection{RSG pulsations for different initial masses}
\label{sec:results:rsg_models}
\begin{figure}[h!]
	\centering 
	\includegraphics[width=0.47\textwidth]{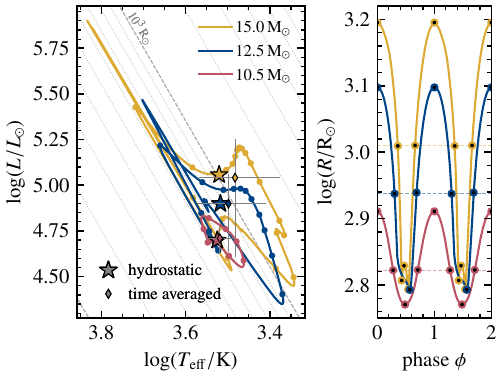}
	\caption{Final hydrodynamic evolution of stars with masses of $10.5$, $12.5$ and $15.0\,\msun$. \emph{Left:} Hertzsprung-Russell diagram of the pulsations. The initial hydrostatic models, as well as the time-averaged luminosity and effective temperature, are indicated. The markers along the pulsation tracks are spaced equally in time every $1/8$, $1/15$, and $1/20$ of the pulsation period for the $10.5$, $12.5$, and $15.0\,\msun$ model, respectively. \emph{Right:} Radius evolution during the pulsations. The dashed horizontal lines show the radii of the hydrostatic models. Markers indicate the times during the pulsation cycle for which we show SN light curves. We define $\phi=0$ at maximal radial extent.}
	\label{fig:HRD_var_m}
\end{figure}
Our models of RSGs experience radial pulsations after core helium burning driven by a $\kappa\gamma$-mechanism for a large luminosity over mass ratio $L/M$ \citepalias{bronner2025}. The resulting variations in luminosity and radius $R$ create a loop in the Hertzprung-Russell diagram (HRD), shown in Fig.~\ref{fig:HRD_var_m}. In general, the more massive the star, the larger are the variations in $R$, $L$, and effective temperature $T_{\rm{eff}}$, and the larger is the pulsation amplitude and the longer its period (Table~\ref{tab:summary-puls}). This is because, to first order, the pulsation period is expected to be proportional to the dynamical timescale $\tau_\mathrm{dyn}$ of the envelope, given by
\begin{equation}
    \tau_\mathrm{dyn} = \sqrt{\frac{R_\mathrm{max}^3}{GM_\mathrm{env}}},
    \label{eq:t_dyn}
\end{equation}
with $M_\mathrm{env}$ the mass of the hydrogen-rich envelope, $R_\mathrm{max}$ the maximum stellar radius during the pulsations, and $G$ the gravitational constant, which increases for higher initial masses. The obtained period is also proportional to the luminosity, leading to a characteristic PL relation for a fixed evolutionary state. We verify that our models are in good agreement with observationally-inferred PL-relations (see Appendix~\ref{app:PL-relations}).

The more massive models also show signs of substructure in the radius evolution and in the HRD track. These originate from the interaction of non-coherently pulsating layers in the envelope and are a result of the thermal restructuring that occurs in these models \citepalias{bronner2025}. Non-sinusoidal light curves with additional substructure have been observed in many variable RSGs \citep[e.g., VX~Sgr,][]{kiss2006a,ren2019a,christodoulou2025a}. Similar substructure can also occur in classical Cepheids \citep[\eg,][]{bono2024a}. The $10.5\,\msun$ model pulsates coherently, producing an almost sinusoidal radius variation and a smooth ellipsoidal track in the HRD.

\subsection{SN light curves and photospheric velocities as a function of initial mass}
\label{sec:results:sn_models}

\begin{figure*}
	\centering
	\includegraphics[width=\textwidth]{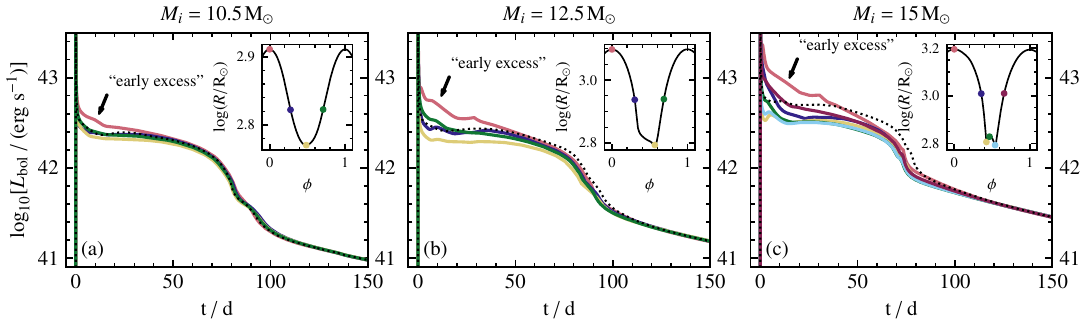}
	\includegraphics[width=\textwidth]{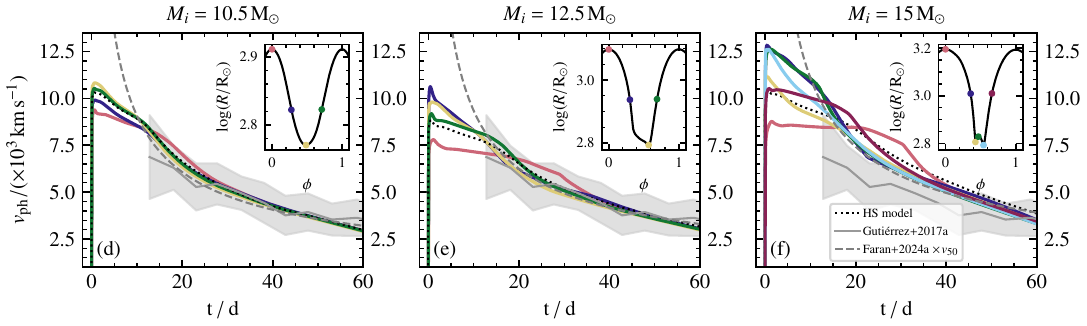}
	\caption{Explosion properties of pulsating RSG progenitors with different initial masses for varying explosion phases (indicated by different colors and highlighted in the insets) as a function of time after shock breakout. The dotted lines indicate the hydrostatic (HS) model. Top panels: bolometric SN light-curves. Bottom panels: qualitative photospheric velocity evolution. The gray line shows the average relation inferred by \citet{gutierrez_2017a} for a sample of 122 observed type II SNe, and the gray region indicates the 1$\sigma$ standard deviation. For comparison, we additionally show the observationally-inferred relation of \citet{faran2014a} scaled by $v_{50}$, the velocity at day 50.}
	\label{fig:LC_vph_var_m}
\end{figure*}

We now characterize the explosion properties of the stellar models. As shown in Fig.~\ref{fig:LC_vph_var_m}, for the higher-mass models, in which the stars experience higher-amplitude pulsations and therefore stronger changes to the outer stellar structure, there are also larger variations in the explosion properties. For the same explosion assumptions, the $15\,\msun$ model shows strong variations in the resulting bolometric SN LCs, ranging from type II-L to type II-P like shapes (see \citetalias{bronner2025} for more detail). For the lower-mass model, we find that the SN LCs vary less and all resemble type II-P supernovae. For larger radii, we observe an early excess emission feature in the SN LC, which is reminiscent of the early emission feature observed in a large fraction of type II SNe and often attributed to CSM \citep[e.g.,][]{morozova_2018}. In our models, the ``early excess" feature arises naturally from the supernova photosphere traveling through the extended envelope. The extended outer atmosphere at these pulsation phases may also act in addition to circumstellar material to cause even larger early luminosity excess and have a similar on the early explosion properties, such as the lower photospheric velocities \citep{moriya2011a}. Thus, what is typically attributed to CSM may in fact be well explained by a low-density extended envelope \citep{falk_arnett_1977_radiation}.

We show the photospheric velocities of SNe at different pulsation phases in the lower panels of Fig.~\ref{fig:LC_vph_var_m}. Just as for the bolometric SN LC, a larger scatter in the photospheric velocity evolution is found for the higher-mass models. They are most visible at early times, where explosions of stars with larger radii lead to lower photospheric velocities.

For SNe with a plateau phase in their SN LCs, we observe that the photospheric velocities converge to similar values during the plateau (from 20d) and follow a similar power-law trend. Comparing with the recent average observed trend of the photospheric velocity of 122 type II supernova by \citet{gutierrez_2017a} that uses the absorption minimum of the FeII $\rm{\lambda} 6159$ line, $v_{\rm{FeII}}$, as a proxy, we find a very good agreement for the lower-mass models (Fig.~\ref{fig:LC_vph_var_m}d and Fig.~\ref{fig:LC_vph_var_m}e). Our $15\,\msun$ model has a higher velocity than the average observed trend, which reflects its higher energy for a similar final mass compared to the lower-mass models. A good agreement is also found with the power-law trend derived for the normalized photospheric velocities of II-P supernovae by \citet{faran2014b}. This trend can be interpreted as a consequence of the phase during which recombination of the hydrogen envelope dominates. At this point, the photosphere follows the outer edge of the recombination zone \citep[e.g.,][]{bersten2011a}, which recedes at a similar rate, independently of the initial explosion. 

The photospheric velocity evolution for the models with larger radii, which have short-plateau or type II-L like SN LCs, deviates from this trend. After reaching the peak, the velocity decreases more slowly than for the models at smaller radii, and stays nearly constant for the models with the largest radii (see Fig.~\ref{fig:LC_vph_var_m}f). Thus, the photospheric velocities at early times (up to about 30d) have the potential to help distinguish between different progenitor structures and to break some of the degeneracies in the explosion properties. This is also the case for non-pulsating models \citep{dessart_2010_determining}, though the photospheric velocities are expected to show smaller variations in these stars, and for a shorter amount of time (within the first 10d, see also Appendix~\ref{app:comp_homology}).
We emphasize that all the photospheric velocity trends described here are qualitative. \texttt{SNEC} is limited to a black-body approximation \citep{morozova2015a}. Accurate predictions for the line velocities, in particular at early and late times during which non-thermal effects play a large role, require radiative-transfer modeling \citep{kozyreva_2020_influence}.
In summary, we find that the observable explosion properties from massive stars that experience a significant re-structuring of their envelope due to their hydrodynamic evolution is qualitatively different compared to commonly-assumed hydrostatic pre-SN models, as their envelopes have a significantly different structure (see also Appendix~\ref{app:comp_homology} for a more detailed comparison).

\subsection{Verifying the progenitor variability of SN~2023ixf and SN~2024ggi}
\label{sec:results:prog_23ixf_24ggi}

Before comparing our models to observational datasets of SNe with observed pulsating progenitors, we verify the reported progenitor variability of SN~20203ixf. Additionally, we do a periodicity study on the progenitor of SN~2024ggi, for which \cite{xiang2024b} report periodic variability, and therefore marking this SN as another potential test case with a pulsating progenitor. 

\begin{figure}
	\centering
	\includegraphics[width=0.47\textwidth]{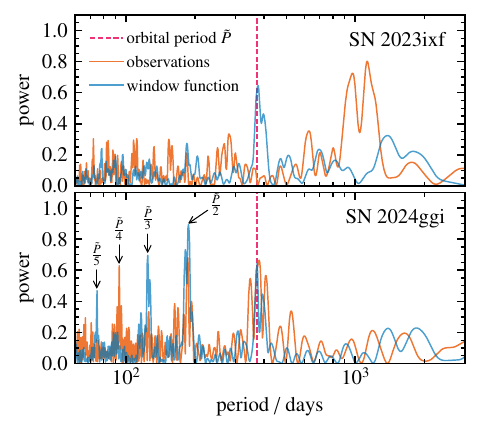}
	\caption{Lomb-Scargle periodogram of the progenitor observations of SN~2023ixf \citep{jencson2023a} and SN~2024ggi \citep{xiang2024b}. The window function of the observations, as well as the orbital period of the Spitzer spacecraft of $P=373.15\,\mathrm{days}$, is shown as a comparison.}
	\label{fig:24ggi_23ixf_LS}
\end{figure}

For SN~2023ixf, \citet{jencson2023a} used archival pre-explosion data from the Spitzer Space Telescope to infer a period of $1119.4^{+132.4}_{-233.3}$ days. We take their photometric data and repeat their analysis using the Lomb-Scargle periodogram from the \texttt{astropy} Python package \citep{astropycollaboration2013a,astropycollaboration2018a,astropycollaboration2022a}. We recover the reported period with two peaks in the periodogram at $975$ and $1125\,\mathrm{days}$ and a full-width-at-half-maximum of $360\,\mathrm{days}$ (Fig.~\ref{fig:24ggi_23ixf_LS}).


For SN~2024ggi, using archival pre-explosion images from the Hubble Space Telescope and the Spitzer Space Telescope, \citet{xiang2024b} found a variable progenitor star with a period of $378.5 \pm 29.4 \, \mathrm{days}$ and attribute this to radial pulsations. Repeating their analysis, we recover the peak in the periodogram at a period of about $380\, \mathrm{days}$, along with several other significant peaks at integer multiples of the base frequency (Fig.~\ref{fig:24ggi_23ixf_LS}). The base period found in the observational data from Spitzer is very close to that of the Spitzer spacecraft of $\sim 373.15\,\mathrm{days}$ \citep{JPLHorizons}. We do not find such a feature in the periodogram of the SN~2023ixf progenitor.

When computing the periodogram of the window function for SN~2024ggi, we find peaks the same periods (Fig~\ref{fig:24ggi_23ixf_LS}). The periodicity may therefore be a result of the periodic observation schedule. More detailed photometric analysis of the progenitor is consequently needed to conclusively identify the progenitor of SN 2024ggi as undergoing radial pulsations. This is beyond the scope of our modeling work. We therefore choose to exclude this object from further analysis.

Recent studies derive an initial mass of $10-15\,\msun$ for SN~2024ggi based on nebular spectroscopy \citep{hueichapan_2025_optical,ferrari_2025_nebular,dessart_2025_SN2024ggi}, $10.5\,\msun$ from the environment close to the explosion site \citep{hong2024a}, and $13\,\msun$ via direct progenitor detection \citep{xiang2024a}. As shown in Sect.~\ref{sec:results:rsg_models}, from our lower-mass models, we would expect that this progenitor is pulsating, probably mildly. The SN~2024ggi progenitor data are unfortunately inconclusive in our analysis. If no evidence for pulsations is found in the RSG progenitor, this would place interesting constraints on the $L/M$ range of the progenitor and also imply that its pre-explosion HRD location may actually be used to estimate its final core mass \citep{temaj2024a}. The unusually long plateau duration of SN~2024ggi \citep{ertini_2025_SN2024ggi}, which implies a large envelope mass, together with a small core mass, points to a likely merger or accretion product \citep{schneider_2025_sn_mergers}. 

In the following, we therefore only consider SN~2023ixf for our model comparison.

\subsection{Model comparison to the progenitor properties of SN~2023ixf}
\label{sec:results:prog_23ixf}
\begin{figure}[h!]
    \centering
    \includegraphics[width=0.47\textwidth]{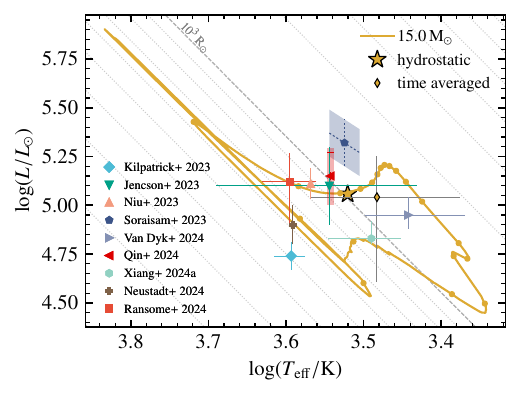}
    \caption{HRD of the $15\,\msun$ RSG model (full line) during one pulsation cycle compared to observationally inferred pre-SN luminosities and temperatures of SN~2023ixf by different studies. Markers on the loop are spaced equally in time every 1/20 of the pulsation period. For reference, we show the location of the hydrostatic model. The time-averaged effective temperature and luminosity of our pulsating stellar model and the associated uncertainties are in good agreement with the spread in observationally inferred values. The shaded parallelogram indicates the possible progenitor location found by \citet{soraisam2023a}. The reported luminosity spread due to the pulsations, reported by \citet{qin2024a}, is shown as a vertical shaded bar, where the luminosity at the inferred explosion time is shown by the horizontal tick.}
    \label{fig:HRD_2023ixf}
\end{figure}

We now compare our models to the properties of SN~2023ixf, beginning with its RSG progenitor.
Several independent studies have attempted to constrain the pre-SN $L$ and $T_{\rm{eff}}$ for SN~2023ixf based on the progenitor data \citep{kilpatrick2023a,jencson2023a,niu2023a,soraisam2023a,vandyk2024a,qin2024a,xiang2024a,neustadt2024a,ransome2024a}. The derived values are shown in Fig.~\ref{fig:HRD_2023ixf}, with the associated uncertainties. Despite being based on similar data, the derived $L$ and $T_{\rm{eff}}$ vary by up to $0.5$~dex, and the inferred initial masses by a factor of 2 \citep{kilpatrick2023a,jencson2023a,soraisam2023a,qin2024a,xiang2024a}.

These findings are based on observations from the Spitzer Space Telescope in Channel~1 ($3.6\,\mu\mathrm{m}$) and Channel~2 ($4.5\,\mu\mathrm{m}$), as well as ground-based near-infrared (NIR) observations in the $J$, $H$, and $K$ bands.

Comparing our model of a $15\,\msun$ pulsating red supergiant, shown by the full line in Fig.~\ref{fig:HRD_2023ixf}, to the inferred progenitor location, we find that the pulsation track in the HRD encloses most observationally derived values within the reported uncertainty ranges. The total spread is also of the same order as the time-averaged spread in $L$ and $T_{\rm{eff}}$ we expect during one pulsation cycle.

Given that the studies we compare to are mostly based on similar data, it appears that the systematic uncertainties are apparently much larger than the reported statistical uncertainties in $L$ and $T_{\rm{eff}}$ (see also  \citealt{davies_2013_temp_rsg,davies2018a,beasor_2025_luminosity_prob}). Different techniques for averaging the stellar LCs may contribute to the spread. Alternatively, the large spread could potentially be understood as a consequence of different studies selecting various parts of the progenitor stellar LC to infer the pre-SN properties, which implies that they would be more sensitive to a certain pulsation phase. However, \citet{qin2024a} take the progenitor pulsations into account and find an amplitude of $0.13\,\mathrm{dex}$ in luminosity assuming a fixed $T_\mathrm{eff}$. Additionally, \citet{soraisam2023a} use a PL relation for their luminosity estimate.
The example of SN~2023ixf thus shows that inferring core masses (let alone initial masses, which are affected by many more uncertainties, including mass loss, binary interactions, and mixing processes like overshooting \citealp{farrell_2020_uncertain,temaj2024a,schneider_2024_preSN_mergers,schneider_2025_sn_mergers}) of exploding RSG from a single pre-SN $L$ and $T_{\rm{eff}}$ is not advisable for RSGs that experience high-amplitude pulsations before explosion. 

An accurate measurement of the progenitor properties shortly before the explosion could constrain the pulsation phase during which the explosion occurred and help break some of the model degeneracies. In the case of SN~2023ixf, this phase is not well constrained, with the last Spitzer observations taken about $1300$~days prior to explosion. There exist NIR observations within $200$~days before explosion \citep{jencson2023a,soraisam2023a}. Unfortunately, the sparse coverage of the progenitor stellar LC in these filters prevents precise constraints on the pulsation phase at explosion. 

\begin{figure}[]
    \centering
    \includegraphics[width=0.5\textwidth]{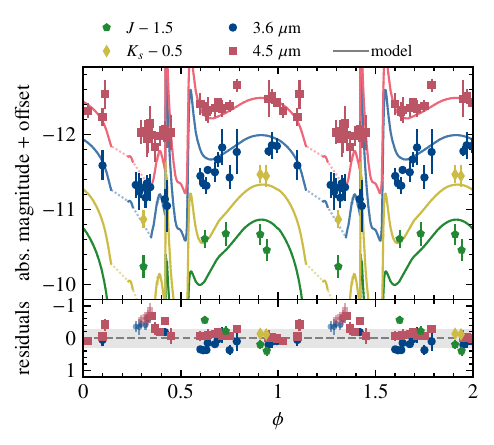}
    \caption{Phase-folded stellar light curve of the $15\,\msun$ pulsating stellar model with a dusty envelope fitted to the observations of the SN~2023ixf progenitor from \citet{jencson2023a}. For the stellar light curve we assume silicate dust \citep{ossenkopf1992a} with $\tau_0=13.5$, $P=1120\,\mathrm{d}$, $\phi_\mathrm{exp}=0.95$ and $T_\mathrm{in} = 1000\,\mathrm{K}$ (see Fig.~\ref{fig:DM_posterior} for full posterior distributions). The shaded-dotted sections of the stellar LCs are linear interpolations because the pulsation model lies outside the MARCS atmosphere models domain (see Appendix~A in \citetalias{bronner2025} for a detailed discussion). The light residual points fall into the interpolated part of the stellar LC and are not considered in the fit.}
    \label{fig:stellar_LC}
\end{figure}

We use DUSTY+MARCS models to derive the stellar LC of the $15\,\msun$ RSG\footnote{The periods of the lower-mass RSG models are incompatible with the progenitor of SN~2023ixf.} with varying dust properties. In Fig.~\ref{fig:stellar_LC}, we fit this model to the observed stellar LC of the SN~2023ixf progenitor from \citet{jencson2023a} (see Appendix~\ref{app:extra-DM-models} for fitting procedure). The pulsation period of our default model ($817\,\mathrm{days}$) is lower compared to the inferred period between around 890 and $1250\,\mathrm{days}$ (see Fig.~\ref{fig:PL_relation})\footnote{Based on this period, we would prefer a more massive progenitor model. However, these models encounter numerical difficulties.}. By re-scaling the stellar LC accordingly (typically stretched by $10-40\,\%$, depending on our assumptions for the dust properties, see Appendix~\ref{app:extra-DM-models}), we find a good agreement between the theoretical predictions and the observations of the SN~2023ixf progenitor. Note that we do not modify the amplitude or the baseline luminosity of the predicted stellar LC. The deviations between the model predictions and the observations are typically less than $0.3$~mag, except for the observations around $\phi=0.4$. Here, the models predict brightness variation of $\sim 2$~mag on timescales of a single day due to ionization effects at the outer boundary. Whether such variations on these short timescales would manifest themselves in stars  with a more advanced treatment of the atmosphere and how these would interact with the dust envelope is not clear. Therefore, we attribute less weight to this phase in the pulsation cycle and accept larger deviations between the models and the observations. The fits in the $J$- and $K_s$-band also show larger deviation compared to the Spitzer-band. This is caused by fewer data points in the filters.

The Spitzer data have coverage of only about $2.5$ periods of the progenitor stellar LC (between $4200$ and $1300$~days before explosion) and do not cover the final year before explosion. This means that the uncertainty of the estimated period is larger than $10\,\%$ \citep{soraisam2023a,kilpatrick2023a,jencson2023a,xiang2024a}. Extrapolating the pulsations to accurately estimate the pulsation phase at explosion is impossible. This causes degeneracies between the fitting parameters of our theoretical model, such as the dust temperature at the inner boundary of the dust envelope or the dust composition, but also the pulsation phase at explosion (see Appendix~\ref{app:extra-DM-models}). Given these constraints, it is not possible to precisely determine the progenitor properties nor the explosion phase $\phi_{\rm{exp}}$ from this dataset. Based on a Bayesian inference approach (see Appendix~\ref{app:extra-DM-models}), we find that values of $\phi_{\rm{exp}}\approx0.9-0.4$, around the maximum radius, are favored (see Fig.~\ref{fig:DM_posterior}). This underlines the importance of pre-explosion stellar LCs for understanding the progenitors of SNe and the last phases of their evolution.

\subsection{Model comparison to the explosion properties of SN~2023ixf}
\label{sec:results:sn_23ixf}
\begin{figure*}[ht!]
	\includegraphics[width=0.5\textwidth]{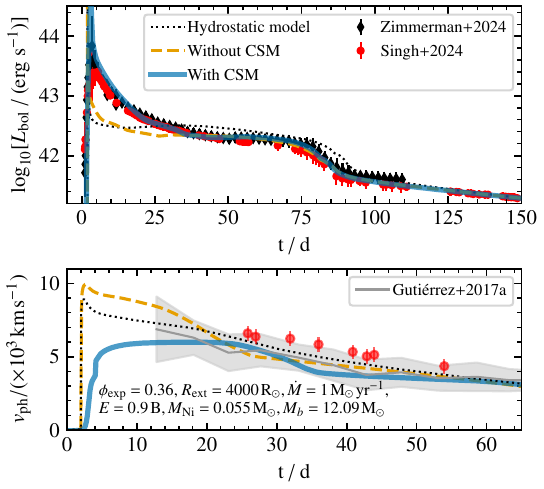}
	\includegraphics[width=0.5\textwidth]{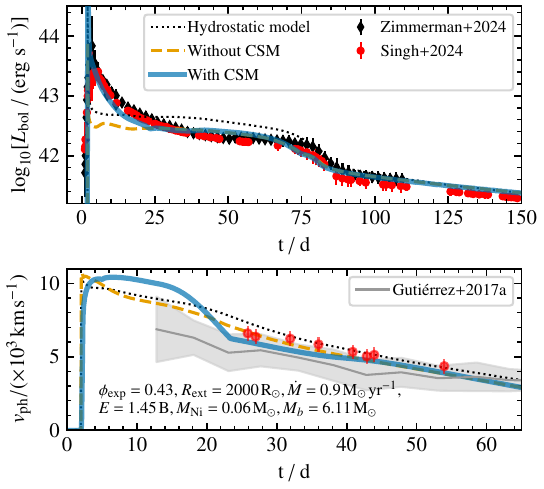}
	\caption{Two example \texttt{SNEC} explosion models based on the $15\,\msun$ RSG model with different radii / explosion phase, and different explosion and CSM properties that both fit very well the observational data of SN2023ixf. For comparison, the hydrostatic models with the same explosion properties without CSM is shown with dotted lines. Top: bolometric SN light-curve. Bottom: As a qualitative comparison, we show the predicted photospheric velocity evolution compared to the average $v_{\rm{FeII}}$ evolution of type-II SNe compiled by \citet{gutierrez_2017a}.}
	\label{fig:fit_LC}
\end{figure*}

Given the overall good agreement between our $15\,\msun$ model and the SN~2023ixf progenitor properties, we focus on this model for inferring the explosion properties, implying a fixed ejecta mass of $M_{\rm{ej}} = 10.7\,\msun$. We note that the $10-40\,\%$ uncertainty in the pulsation period of our model compared to the SN~2023ixf progenitor period (see Sect.~\ref{sec:results:prog_23ixf_24ggi}) implies an underlying uncertainty in this ejecta mass after explosion. We construct 2016 \texttt{SNEC} models in which we vary the explosion properties ($E$ from 0.9 to 1.45B, $M_{\rm{Ni}}$ from $0.05$ to $0.06\,\msun$, $M_{\rm{b}}$ from $0.5$ to $0.99\, M$) for the range of explosion phases $\phi_{\rm{exp}}$ shown in Fig.\ref{fig:LC_vph_var_m}c.
The parameter space of explosion properties is affected by numerous well-known degeneracies that make it difficult to further constrain $\phi_{\rm{exp}}$ \citep{dessart_2019_difficulty,martinez_2019_mass,goldberg2019a,goldberg_2020_value,dessart_2023_using, fang_2025_diversity}. For instance, both lower explosion energies and smaller pulsation amplitudes lead to lower explosion luminosities and longer plateau durations (see also Appendix~\ref{app:comp_homology}). Two example SN LCs and photospheric velocities based on progenitors at explosion phases with significantly different radii ($R_{*}=1009\,\rsun$ for $\phi_{\rm{exp}}=0.36$ and $R_{*}=643\,\rsun$ for $\phi_{\rm{exp}}=0.43$) and explosion energies ($E=0.9\,\rm{B}$ and $E=1.45\,\rm{B}$, respectively) that fit the late-time ($> 25\,\rm{d}$) dataset very well are shown in Fig.~\ref{fig:fit_LC}.
To obtain a good fit for the first part of the SN LC, an additional component is needed. This is supported independently by the spectral evolution of SN~2023ixf, which is consistent with interaction with CSM \citep[\eg][]{jencson2023a,jacobson-galan_2023_core-collapse,smith2023,zimmerman2024a,singh2024}. We construct a CSM model following the approach of \citet{morozova_2018}. The CSM is computed based on a steady-state wind-like density profile $\rho_{\rm{CSM}}$ with a velocity law of $\beta=0$ \citep{moriya2011a,moriya2023}:
\begin{equation}
	\rho_{\rm{CSM}}(r) = \frac{\dot{M}}{4\pi v_{w}r^{2}},
\end{equation}

where $\dot{M}$ is the mass loss rate, $v_w$ the wind velocity, and $R_{\rm{ext}}$ the radial extent of the CSM \citep{moriya2011a,moriya2023}. Based on observational constraints, we set the terminal wind velocity to $v_{w, \rm{inf}}=115 \,\rm{km}\rm{s}^{-1}$ \citep{smith2023}. As shown in Fig.~\ref{fig:fit_LC}, the data constraints are not sufficient to discriminate between different CSM models, explosion energy, or pulsation phases, as we obtain very good SN LC fits for two models with different CSM properties ($m_{\rm{CSM}}=0.43\,\msun$ for $\phi_{\rm{exp}}=0.36$ and $m_{\rm{CSM}}=0.23\,\msun$ for $\phi_{\rm{exp}}=0.43$, and a factor two difference in $R_{\rm{ext}}$). Measurements of photospheric velocities at earlier times would provide clearer constraints. Given the black-body approximation of \texttt{SNEC}, it is not the appropriate tool for deriving such constraints, which would require radiation-transfer calculations.

Exploring a large range of SN explosion parameters (see above) and CSM properties ($\dot{M}$ from $0.2$ to $0.9\,\msun\,\rm{yr}^{-1}$ and $R_{\rm{ext}}$ from $500$ to $6000\,\rsun$), we tentatively exclude explosions at maximum radial extent, as these result in rapidly declining SN LCs resembling type II-L SNe that do not fit the SN LC of SN~2023ixf well (see Fig.~\ref{fig:LC_vph_var_m}). However, we do not find strong constraints on the explosion phase, given the large degeneracies in the SN properties and the presence of CSM.

In summary, combining our results for the pre-SN and explosion properties of SN~2023ixf, we find that our $15\,\msun$ fits the data very well. Our progenitor modeling favors a $\phi_{\rm{exp}}$ near radius maximum, but we do not find strong constraints from the SN LC modelling due to underlying degeneracies. As pointed out by \citet{hsu_2024_one}, the pulsation period of SN~2023ixf implies an independent constraint to the ejecta/envelope mass - radius - explosion energy relation obtained from the plateau phase of type II SN LCs \citep{popov1993a,kasen_2009_typeII,goldberg2019a} and can therefore help break these degeneracies. The ejecta mass of $M_{\rm{ej}} = 10.7\,\msun$ is in a similar range, but slightly higher than that obtained by several studies that performed modeling of the plateau based on hydrostatic models, where $M_{\rm{ej}} = 8 - 9.5\,\msun$  \citep{hiramatsu_2023_SN2023ixf,singh2024,hsu_2024_one,bersten_2024_progenitor,vinko_2025_SN2023ixf}. However, as mentioned in Section~\ref{sec:results:prog_23ixf}, there is an underlying uncertainty in this ejecta mass due to the 10-40\% uncertainty in the pulsation period of the progenitor used to obtain this model.
If $\phi_{\rm{exp}}$ (and therefore the pre-explosion radius) were to be better constrained, we could obtain a full solution for the explosion properties of SN~2023ixf.

\section{Discussion}
\label{sec:discussion}

\subsection{RSG pulsations}
The mass range at which radial RSG pulsations become important and significantly affect the pre-explosion structure sensitively depends on our assumptions regarding the envelope physics, in particular convection. In RSGs, both observations \citep[\eg][]{montarges_2017_convective} and expensive 3D modeling efforts \citep{goldberg2022a,ma_2024_Betelgeuse} have shown that convection is an intrinsically 3D process, where large-scale convective cells easily break the spherical symmetry. Convection in these RSGs happens on similar timescales as the pulsations \citepalias{bronner2025} and it is unclear how exactly they will affect one another, but first modeling efforts find that these effects will probably be minor \citep{heger1997a,langer1971a}. Latest 3D-RSGs models suggest that the fundamental radial mode dominates the pulsation \citep{ma2025b}, in line with observational expectations \citep[e.g.,][]{kiss2006a,ren2019a}.

One-dimensional models typically simulate convection using mixing-length theory (MLT), where the efficiency is determined by the mixing length parameter $\alpha_{\rm{MLT}}$.  Its value is uncertain \citep{joyce_2020_standing}, but recent 3D models of RSGs find that it may be in the range of $\alpha_{\rm{MLT}} \approx 1.4 - 3$ throughout the envelope \citep{goldberg2022a,ma2025b}. Observationally, this parameter is constrained by the observed radii of RSGs, which range from $400\,\rsun$ to $1400\,\rsun$ \citep{davies_2013_temp_rsg,montarges_2014_properties,o_gorman_2015_VYCan} and translate into a typical RSG location on the HRD. For example \citet{ekstroem_2012_grids} use $\alpha_{\rm{MLT}}=1.6$ to fit observations by \citet{levesque_2005_effective}, while \citet{chun2018a} find a metallicity-dependent $\alpha_{\rm{MLT}}$ value between 1.8 and 2.9. Recent type-II SNe studies (based on hydrostatic stellar models) fit the color evolution better with smaller RSG radii of $\approx 500\,\rsun$ \citep{dessart2013a}, corresponding to $\alpha_{\rm{MLT}}\approx 2.5$. Choosing a larger $\alpha_{\rm{MLT}}$ value, as in \citet{goldberg2020a}, would shift the occurrence of high-amplitude pulsations to higher-mass progenitors. We note that the use of MESA's \texttt{MLT++} scheme in our models \citepalias[see][]{bronner2025} reduces this apparent difference, as it leads to more compact RSGs.

In general, the periods obtained from the hydrodynamic modeling of the RSGs match the expectations from observational and theoretical PL relation (see Appendix~\ref{app:PL-relations}). The predicted amplitudes are, however, larger than observed amplitudes \citep[e.g.,][]{ren2019a}. This can be explained by the observational bias towards core helium-burning RSGs, which are longer lived than RSGs in more advanced burning phases. Additionally, simplified modeling of convection via MLT and its coupling to the pulsations can directly alter the modeled amplitudes, as well as the choice of the boundary conditions. For a more detailed discussion of these effects, see Sect.~5.4 in \citetalias{bronner2025}, but also \citet{joyce2020a} and \citet{suzuki2025a}.
 
Another uncertain factor is the mass lost by the star before the explosion. If, for any reason, the star loses mass before explosion, the $L/M$ ratio will increase, and so will the pulsation amplitude. The effect described here may thus only apply to higher-mass RSGs or to stars that lose their envelopes before explosion, for example, due to binary mass transfer. 

Since most massive stars live in interacting binaries \citep{sana_2012_binary}, their pre-SN structures will be different compared to the single stars described here \citep{podsiadlowski1992a,laplace_2021_different,schneider2021a,schneider_2024_preSN_mergers}, affecting the amplitude of RSG pulsations and the mass range at which they occur. For example, the signs of large-scale asymmetries in SN~2023ixf have been interpreted as signatures of past binary interactions \citep[\eg][]{singh2024,hsu_2024_one}, which may imply more mass loss before explosion compared to single stars. We expect that not only donors in binary systems, but also accretors and mergers may experience such pulsations, based on their pre-SN structures \citep{schneider_2024_preSN_mergers,schneider_2025_sn_mergers}. The interplay between envelope pulsations and binary interactions may further affect the pre-SN structure and is another uncertain and important aspect that we will study in the future.  

\subsection{Circumstellar material}
Our model does not self-consistently eject material due to the pulsations. However, the development of sub-surface shocks and the low binding energy at the phase of radius minimum (see Appendix~\ref{app:props}  and \citetalias{bronner2025}) lead us to speculate that envelope ejections shortly before core collapse may be likely in pulsating SN progenitors \citep[see also][]{clayton2018a}, naturally producing the CSM observed in SN~2023ixf. Pulsation-driven mass ejections are expected for higher-mass progenitors \citep{clayton2018a}, may explain the origin of this CSM \citep{yoon2010a}, and could be observed as SN precursors in case of particularly high-mass ejections. Alternatively, the thermal restructuring phase we identify in our models for high amplitude pulsation can also radiate away about $5\times10^{46}\,\rm{erg}$ of energy over about 30d (similar to observed luminous red novae, see \citetalias{bronner2025}), and could also be observed as a SN precursor. The existence of a SN precursor was recently proposed to explain the early explosion properties of SN~2023ixf \citep{kozyreva_2025_SN2023ixf}. Based on our model it would be unlikely that such a precursor would be detected so shortly before explosion. Based on archival searches, no bright precursor is detected for SN~2023ixf in the past 5 years before explosion \citep{dong_2023_comprehensive}, which would disfavor a single episode of high mass ejection and would instead be consistent with mass loss over a long period of time, possibly due to pulsations. 

From the pulsation period of the SN~2023ixf, we would expect a higher-mass progenitor than the one modeled here, which would be more likely to experience mass loss due to pulsations. Very recently, \citet{sengupta_2025_dance} showed that the pulsation-driven mass loss, described following the prescriptions by \citet{yoon2010a} and \citet{clayton2018a}, can lead to significant mass loss in RSGs and explain the CSM structure of SN~2023ixf reasonably well. For higher-mass progenitors with stronger mass loss, this could lead to a runaway mass-loss process of pulsations with increased amplitude, triggering stronger mass ejections until the star has lost so much mass that it no longer appears as an RSG \citepalias{bronner2025}. This may explain the lack of high-luminosity RSG SN progenitors (also known as the missing RSG problem, \citealt{smartt2009b,smartt2015a}).\\
We emphasize that not all signatures of CSM in SN explosions are necessarily connected to pre-SN mass loss. Instead, ``early excess'' features identified in a large fraction of type-II SN light curves can alternatively be explained by extended progenitors \citep[\eg][]{falk_arnett_1977_radiation,ercolino2024a}. Here, we show that radial pulsations, in particular in high-mass RSGs, can naturally lead to extended SN progenitors. We predict that these signatures should be more prevalent in SNe from progenitors that experience higher-amplitude pulsations. Recently, \citet{fuller2024a} discussed the effect of RSGs chromospheres on SNe LCs and showed that these can translate into early luminosity excess. Taking these chromospheres, or similarly, ``effervescent zones" that may be created by a combination of pulsation-driven outflows and convection in the envelope \citep{soker_2021_effervescent,soker_2023_effervescent_23ixf} into account for our pulsating progenitors would thus further enhance the early emission we described. We expect that together with radial pulsations, this could also help explain the prevalence of early ``flash ionization" signatures in type II SNe \citep{bruch_2021_large}, and their long rise time until shock breakout \citep{foerster2018}, without the need of invoking unknown pre-SN mass-loss processes. Recently-reported observations show that type-II SNe with early ionization signatures have typically luminous fast-declining SN LCs \citep{jacobson-galan_2025_finalIII}. This is consistent with high mass stars preferentially exploding with large radii due to pulsations \citepalias{bronner2025}.  Pulsations can also explain the ~0.2 dex spread between the i-band and spectral energy distribution luminosities in observed RSGs \citep{beasor_2025_luminosity_prob}.

\section{Conclusion}
\label{sec:conclusion}

We investigate the pre-SN evolution of massive RSG between $10.5$ and $15\,\msun$, focusing on their hydrodynamical evolution and the effect on their explosion properties. In line with previous studies \citep{heger1997a,yoon2010a,clayton2018a,goldberg2020a}, we find that they all develop radial envelope pulsations. In our more massive progenitors, the pulsations lead to a re-structuring of the envelope that leads to a significantly different pre-SN envelope structure compared to hydrostatic models. We show that pulsations significantly affect the pre-SN properties of RSGs, including their location on the HRD, and that their importance increases with mass. Based on our findings, we caution against inferring core masses (let alone initial masses) for massive pulsating RSGs from a single pre-SN HRD location.
 
Studying the explosion properties of these pulsating RSGs, we show that the changing structure as a function of pulsation phase can lead to significant variations compared to hydrostatic progenitors. In particular, extended progenitors appear to have an ``early excess" in their SN LCs, even for our low-mass progenitors. This early excess is observed in a large fraction of SNe (up to 60\%, \citealt{morozova_2018}) and is often attributed to CSM of unknown origin \citep{morozova_2018,hinds_2025_inferring}. Here we show that it can at least in part be the result of a more extended progenitor structure due to radial pulsations, and could potentially be extended further if chromospheres of RSGs were taken into account \citep{fuller2024a}. This same effect could also provide a natural explanation for ``flash ionization" signatures in the spectra of a large fraction of type-II SNe \citep{bruch_2021_large}, which are preferentially found in luminous SNe with fast-declining SN LCs \citep{jacobson-galan_2025_finalIII}. Analytical scaling relations typically used for inferring SN properties \citep{popov1993a,kasen_2009_typeII,goldberg2019a} should therefore be used with caution, in particular for luminous supernovae with early excess (see Appendix~\ref{app:comp_homology}).

As a function of the pulsation phase, we find significant variations in the photospheric velocity evolution, in particular at early times (before $25\,\rm{d}$), whose importance increases for higher mass progenitors that experience larger-amplitude pulsations. These features can help break long-standing degeneracies \citep{kasen_2009_typeII,goldberg2019a} in the explosion properties of type II SNe. 

We apply our models to the recent SN~2023ixf and SN~2024ggi, for which pulsating RSG progenitors were reported. We find insufficient evidence for the claimed pulsation period of the SN~2024ggi progenitor, as it matches Spitzer’s orbital period. Based on its explosion properties, we speculate that it may be a binary stellar merger product.

We recover the reported pulsation period of SN~2023ixf and demonstrate that the pre- and post-explosion characteristics can all be fit very well with our exploding pulsating RSG models. However, the data coverage is insufficient to break the degeneracies in explosion parameters and pulsation phase. Based on the good agreement between our model and the pre-SN and explosion parameters, we tentatively constrain the explosion phase to be at intermediate pulsation radii. If it were precisely determined, we could find a full solution for the explosion properties of SN~2023ixf. This shows the large potential of leveraging pulsating progenitors for constraining SN properties and underlines the limitation of hydrostatic models. We speculate that the CSM in SN~2023ixf is also connected to pulsations through pulsation-driven outflows \citep{yoon2010a,clayton2018a,sengupta_2025_dance}. Such outflows are expected to be more prevalent in higher-mass progenitors due to the dependence of the $\kappa\gamma$-mechanism on the $L/M$ ratio \citepalias{bronner2025}.

The exact mass range in which pulsations play an increased role for the pre-SN structure is subject to large uncertainties regarding the mixing processes in stars, their mass loss, and their binary interactions (see also \citetalias{bronner2025}). However, in any case, we expect that RSGs on the more massive end experience large-amplitude radial pulsations shortly before their explosion.

This study emphasizes the importance of the hydrodynamic evolution of stars, in particular from $\sim15\,\msun$, for their last evolutionary stages, their pre-SN, and explosion properties. With the start of next-generation all-sky surveys such as the Legacy Survey of Space and Time of the Vera Rubin Telescope, it will become easier to detect SNe with variable progenitors, giving us more opportunities to further understand the importance and prevalence of pulsations in SN progenitors.

\section*{Acknowledgments}
We thank the anonymous referee for their thorough review, which helped improve this work. We thank Avinash Singh and Brian Hsu for generously providing early datasets of SN~2023ixf. We also thank Takashi Moriya, Jared Golberg, Ebraheem Farag, and Emma Beasor for constructive discussion that helped improve this work.
VAB, EL, FRNS and PhP acknowledge support from the Klaus Tschira Foundation.
This work has received funding from the European Research Council (ERC) under the European Union’s Horizon 2020 research and innovation programme (Grant agreement No.\ 945806), and is supported by the Deutsche Forschungsgemeinschaft (DFG, German Research Foundation) under Germany’s Excellence Strategy EXC 2181/1-390900948 (the Heidelberg STRUCTURES Excellence Cluster). VAB acknowledges support from the International Max Planck Research School for Astronomy and Cosmic Physics at the University of Heidelberg (IMPRS-HD). EL acknowledges support through a start-up grant from the Internal Funds KU Leuven (STG/24/073) and through a Veni grant (VI.Veni.232.205) from the Netherlands Organization for Scientific Research (NWO). This research was supported by the Munich Institute for Astro-, Particle and BioPhysics (MIAPbP) which is funded by the Deutsche Forschungsgemeinschaft (DFG, German Research Foundation) under Germany's Excellence Strategy -- EXC-2094 -- 390783311.
\software{
	Astropy \citep{astropycollaboration2013a,astropycollaboration2018a,astropycollaboration2022a}, 
	CMasher \citep{velden2020a}, 
	Matplotlib \citep{hunter2007a}, 
	NumPy, \citep{harris2020a}, 
	PyAstronomy \citep{czesla2019a}, 
	PYPHOT \citep{fouesneau2025a}, 
	SciPy \citep{virtanen2020a}
}

\appendix

\section{Additional model properties}
\label{app:props}
For reference, we show additional properties of our models in Tables~\ref{tab:summary-exp} and \ref{tab:summary-puls}. The assumed explosion parameters for our default \texttt{SNEC} models are obtained through the semi-analytical neutrino-driven explosion code of \citet{muller2016a}, with calibrations as in \citet{schneider2021a}, \citet{temaj2024a}, and \citet{laplace_2025_written}.

\begin{deluxetable}{ccccc}
	\tablewidth{0pt}
	\tablecolumns{5}
	\caption{Summary of the explosion properties of the RSG models with different initial mass $M_\mathrm{i}$. The explosion parameters are determined with the \citet{muller2016a} model with calibrations as in \citet{schneider2021a}.\label{tab:summary-exp}}
	\tablehead{
		\colhead{$M_\mathrm{i}$} & \colhead{$E$} & \colhead{$M_\mathrm{NS}$} & \colhead{$M_\mathrm{Ni}$} & \colhead{$M_b$} \\
		\colhead{$[\msun]$} & \colhead{$10^{51}\,\mathrm{erg}$} & \colhead{$[\msun]$} & \colhead{$[\msun]$} & \colhead{$[\msun]$}
	}
	\startdata
	10.5 & 0.916 & 1.387 & 0.0295 & 8.70 \\
	12.5 & 0.871 & 1.597 & 0.0450 & 9.88 \\
	15.0 & 1.60  & 1.502 & 0.0863 & 11.0 \\
	\enddata    
\end{deluxetable}

We summarize the pulsation properties of the model in Table~\ref{tab:summary-puls}. With increasing mass, the amplitudes in the stellar radius $R$, luminosity $L$, and effective temperature $T_\mathrm{eff}$ become larger. To first order, the pulsation period scales linearly with the dynamical timescale $\tau_\mathrm{dyn}$ of the model. We find that the ratio of surface velocity and escape velocity $v_\mathrm{surf}/v_\mathrm{esc}$ also increases with mass. This is expected because the pulsations become stronger for larger $L/M$. For $v_\mathrm{surf}/v_\mathrm{esc} > 1$, we would expect dynamical mass ejection. This agrees with the results in \citet{clayton2018a} where dynamical mass ejections only occur for $\log(\frac{L/\lsun}{M/\msun}) > 4.1$. We expect the mass ejection to happen typically shortly after maximum compression (around $\phi\sim 0.65$), where $v_\mathrm{surf}/v_\mathrm{esc}$ reaches its maximum value (see also \citealt{clayton2017a} and \citealt{clayton2018a}).

We use \texttt{GYRE} \citep{townsend2013a} to determine the period of the fundamental radial mode, $P_\mathrm{GYRE}$, based on the initial hydrostatic models \citep[see also][]{joyce_2020_standing}. Additionally, we computed the pulsation period, $P_\mathrm{growth}$, during the exponentially growing phase of the pulsations (see Sect.~3.1 in \citetalias{bronner2025}). We find that $P_\mathrm{GYRE}$ and $P_\mathrm{growth}$ are compatible within the uncertainties, but the period $P$ of the steady-state pulsation deviates. This is expected because, after the initial transient phase and potentially a catastrophic cooling event, the envelope structure changes, and therefore, its dynamical properties also change. This also means that it is not possible to accurately determine the pulsation period using a perturbation analysis of the initial hydrostatic model.

\begin{deluxetable*}{cccccccccccc}
	\tablewidth{0pt}
	\tablecolumns{12}
	\caption{Summary of the pulsation properties of the RSG models with different initial mass $M_\mathrm{i}$ and $L/M$ (in solar units). We show peak-to-peak amplitudes of the radius~$R$, luminosity~$L$, and effective temperature~$T_\mathrm{eff}$, the pulsation period in units of the dynamical timescale (Eq.~\ref{eq:t_dyn}), the maximum of the surface velocity $v_\mathrm{surf}$ over the escape velocity $v_\mathrm{esc}$, with $v_\mathrm{esc} = (2GM/R)^{1/2}$, and the corresponding pulsation phase.\label{tab:summary-puls}}
	\tablehead{
		\colhead{$M_\mathrm{i}$} & \colhead{$\log(L/M)$} &  \colhead{$P_\mathrm{GYRE}$} & \colhead{$P_\mathrm{growth}$} & \colhead{$P$} & \colhead{$\Delta \log R$} & \colhead{$\Delta \log L$} & \colhead{$\Delta \log T_\mathrm{eff}$} & \colhead{$M_\mathrm{env}$} & \colhead{$P/\tau_\mathrm{dyn}$} &\colhead{$\bigl(\frac{v_\mathrm{surf}}{v_\mathrm{esc}}\bigr)_\mathrm{max}$} & \colhead{$\phi_{v_\mathrm{surf,max}}$} \\
		\colhead{$[\msun]$} & \colhead{} & \colhead{[days]} & \colhead{[days]} & \colhead{[days]} & \colhead{[dex]} & \colhead{[dex]} & \colhead{[dex]} & \colhead{$[\msun]$} & \colhead{} & \colhead{} & \colhead{}
	}
	\startdata
	10.5 & 3.71 & 470  & $477^{+13}_{-15}$  & $490^{+5}_{-7}$   & 0.14 & 0.28 & 0.10 & 6.56 & 2.92 & 0.180 & 0.76\\
	12.5 & 3.86 & 745  & $740^{+40}_{-40}$  & $702^{+8}_{-7}$   & 0.31 & 1.12 & 0.33 & 6.97 & 2.27 & 0.430 & 0.67\\
	15.0 & 3.91 & 1010 & $920^{+110}_{-70}$ & $817^{+8}_{-11}$  & 0.40 & 1.45 & 0.49 & 6.99 & 1.86 & 0.507 & 0.62\\
	\enddata    
\end{deluxetable*}

\section{Comparison to observed and predicted PL-relations of RSGs}
\label{app:PL-relations}

\begin{figure}
    \centering
    \includegraphics[width=0.47\textwidth]{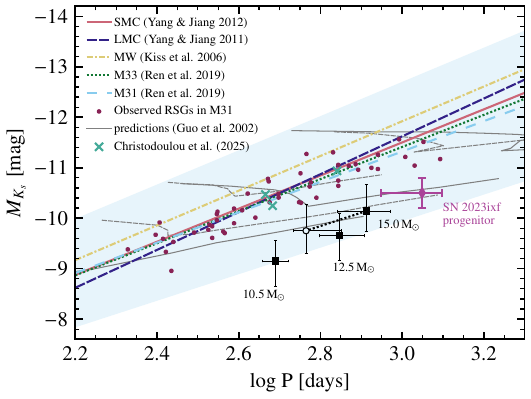}
    \caption{Comparing models of pulsating RSGs to observationally determined PL relations in various environments \citep{kiss2006a,yang2011a,yang2012a,ren2019a}. The shaded region shows the uncertainty of the PL relation for M31 and is characteristic for all the others. Theoretical predictions are from \citet{guo2002a} for evolving 15, 20, and $30\,\msun$ RSGs at $Z=0.02$ and 0.01 (full and dashed lines respectively), adjusted with bolometric corrections from \citet{josselin2000a}. For comparison, we show the recent observations from \citet{christodoulou2025a}. The SN~2023ixf progenitor is taken from \citet{jencson2023a}. The model with an open symbol connected to the $15\,\msun$ RSG is taken at the end of core helium burning.}
    \label{fig:PL_relation}
\end{figure}

We compare the periods of the steady-state pulsations obtained from our RSG models with observationally determined PL relations recovered in various environments (Fig.~\ref{fig:PL_relation}). For our three RSG models, the uncertainty the pulsation periods is determined by using the same sampling as in the Spitzer observations of the SN~2023ixf progenitor. The uncertainty of the $K_s$-band magnitude takes into account different dust properties of the \texttt{DUSTY} radiation transfer models. As shown in Fig.~\ref{fig:PL_relation}, there is a general trend of our models being located below the observational PL relations from large samples of RSGs. This is expected from theoretical predictions (full and dashed black lines in Fig.~\ref{fig:PL_relation}), as RSGs generally evolve to higher luminosities and longer pulsation periods at later evolutionary stages. In particular, the prediction for a $Z=0.02$ $15\,\msun$ star close to core-collapse is very close to our $15\,\msun$ RSG model (black point), for which we also find a similar evolution compared to its pulsation period at the end of core He burning (see open symbol and dotted line in Fig.~\ref{fig:PL_relation}). Observational samples of RSGs are dominated by stars in their long-lived core-He-burning phase, and are thus located around lower base periods. The observed periods of the progenitor of SN~2023ixf (purple point on Fig.~\ref{fig:PL_relation}) cover a wide range between around 890 and $1250\,\mathrm{days}$ follows the same trend as our model, and is compatible with our $15\,\msun$ model. In summary, we verify that our models are compatible both with observationally-inferred and theoretically-predicted PL relations for RSGs.

\section{Fitting progenitor light curve of SN~\lowercase{2023ixf}}\label{app:extra-DM-models}

The progenitor stellar LC are computed based on MARCS atmosphere models \citep{gustafsson2008a} with the DUSTY radiation transfer code \citep{ivezic1997a,ivezic1999a} and depend on the total optical depth $\tau_0$ (see Section~2.3 in \citetalias{bronner2025}). We consider three different dust compositions: warm silicates based on \citet{ossenkopf1992a}, silicates based on \citet{draine1984a}, and graphite based on \citet{draine1984a}. The dust temperature at the inner boundary $T_\mathrm{in}$ is varied between $800$, $1000$, and $1200\,\mathrm{K}$. The computed stellar LCs are stretched and shifted to match any pulsation period $P$ and pulsation phase at explosion $\phi_\mathrm{exp}$.

We use a Bayesian inference approach to fit the progenitor stellar LC to the observed data collected in \citet{jencson2023a}, with inferred quantities $\vec{\theta} = (\tau_0, P, \phi_\mathrm{exp})$. We choose a uniform prior for $\tau_0$ with bounds between $0.5$ and $25$. The prior for the pulsation period is proportional to the Lomb-Scargle power (Fig.~\ref{fig:24ggi_23ixf_LS}), bounded between $500$ and $1500\,\mathrm{d}$. The explosion phase $\phi_\mathrm{exp}$ is treated as a circular variable with a uniform prior between 0 and 1. To compare our model to the data, we choose a Gaussian likelihood function,
\begin{equation}
	\ln \mathcal{L}(\vec{\theta}) = -\frac{1}{2}\sum_i^{N_\mathrm{obs}} \left[\frac{\left(m(\vec{\theta}, t_{i}) - m_{\mathrm{obs},i}\right)^2}{s_i^2} + \ln(2\pi s_i^2)\right],
\end{equation}
where $m(\vec{\theta}, t)$ is the modeled magnitude at time t, $m_{\mathrm{obs},i}$ is the observed magnitude at time $t_i$, $s_i^2 = \sigma_i^2 + f^2 m(\vec{\theta}, t_i)$, and $\sigma_i$ is the observational uncertainty. The variable $f$ accounts for underestimated uncertainties of the posterior distribution, and we treat it as another inference quantity with a log-uniform prior \citep{hogg2010a}. Some data of \citet{jencson2023a} have very low uncertainties ($\sim 0.04\,\mathrm{mag}$). We set an additional lower limit of $0.1\,\mathrm{mag}$ to the uncertainties to avoid underestimating the errors. The likelihood function sums all observations $N_\mathrm{obs}$ in $J$-band, $K_s$-band, Spitzer channel~1 ($3.6\,\mu\mathrm{m}$) and channel~2 ($4.5\,\mu\mathrm{m}$) reported in \citet{jencson2023a}. If an observation falls into the phase range where the stellar LC cannot be predicted, this observation is not taken into account. We sample from the posterior distribution using the \texttt{emcee} Python package \citep{foreman-mackey2013a}, which is an implementation of the affine invariant Markov chain Monte Carlo ensemble sampler from \citet{goodman2010a}.

\begin{figure}
    \centering
    \includegraphics[width=0.47\textwidth]{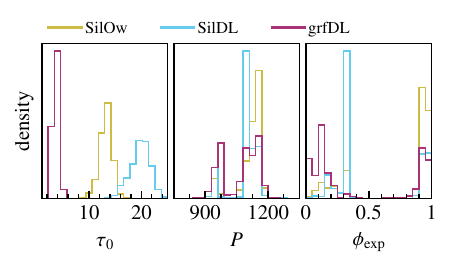}
    \caption{Posterior distributions of the optical depth $\tau_0$, the pulsation period $P$, and the pulsation phase at explosion $\phi_\mathrm{exp}$. Posteriors are shown for the different dust compositions: warm silicates based on \citet{ossenkopf1992a} (SilOw), silicates based on \citet{draine1984a} (SilDL), and graphite based on \citet{draine1984a} (grfDL). The posteriors are summed over all three possible values for $T_\mathrm{in}$.}
    \label{fig:DM_posterior}
\end{figure}

\begin{figure}
    \centering
    \includegraphics[width=0.47\textwidth]{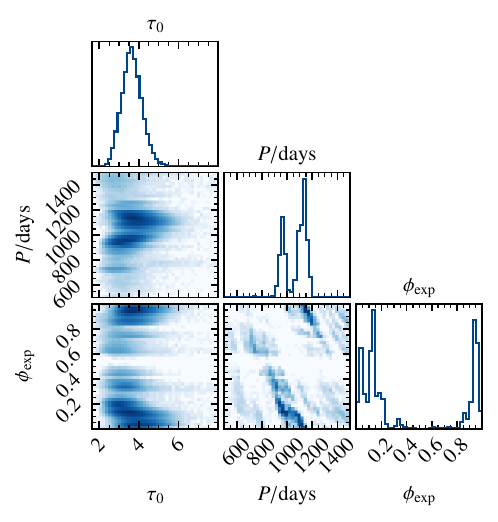}
    \caption{Corner plot showing the posterior distribution assuming graphite dust \citep{draine1984a}. The 1D histograms use a linear ordinate while the 2D histograms using a logarithmic scale on the coloring.}
    \label{fig:DM_posterior_corner}
\end{figure}

The marginalized posterior distributions for $\tau_0$, $P$, and $\phi_\mathrm{exp}$ are shown in Fig.~\ref{fig:DM_posterior}, where we have integrated over all values of $T_\mathrm{in}$ for a given dust composition. The dust composition directly sets the mode of the $\tau_0$-distribution, where graphite dust requires the lowest values ($\tau_0~\sim 4$) and silicates require the highest values ($\tau_0 \sim 14-20$). The dust composition has only a small impact on the posterior distribution of the period $P$. This is expected because the period prior, which is based on the Lomb-Scargle power, is already very informative and uses the observed data. In a Bayesian view, the power can also be seen as the posterior distribution when fitting a sinusoidal model to the data \citep{jaynes1987a}. The posterior of the explosion phase $\phi_\mathrm{exp}$ shows a larger dependence on the dust composition. Taking all dust compositions into account, we find that $\phi_\mathrm{exp}$ can be constrained between $0.9$ and $0.4$, i.e., around the maximum radius of the progenitor. This phase is much longer lived than the compressed phase because of a longer dynamical timescale (see Fig.~\ref{fig:HRD_var_m}).

We show a corner plot of the posterior distribution for graphite based dust (grdDL) in Fig.~\ref{fig:DM_posterior_corner}. There is a clear correlation between the inferred pulsation period P and the pulsation-phase at explosion $\phi_\mathrm{exp}$. Nonetheless, there is a clear gap in the $\phi_\mathrm{exp}$ distributions around $0.5$.

\section{Comparison of SN light curves from hydrostatic and hydrodynamic progenitors}
\label{app:comp_homology}

\begin{figure}
	\centering
	\includegraphics[width=0.5\textwidth]{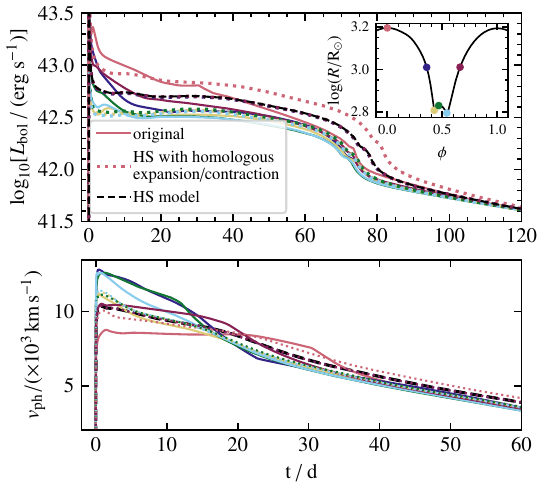}
	\caption{Supernova light curve (top) and photospheric velocity evolution (bottom) of the $15\,\msun$ stellar model. Here we compare the explosion outcome of the hydrodynamic pre-SN model to the hydrostatic pre-SN model, and to pre-SN models that assume homologous contaction/expansion of the hydrostatic model to the same radius as the one reached by the hydrodynamic model for different pulsation phases shown in the inset. All models have the same explosion energy, ejecta mass, and mass and distribution of $^{56}\rm{Ni}$.}
	\label{fig:comp_SNLC_homology}
\end{figure}
Here, we investigate the difference in the supernova properties obtained between our approach of modelling the hydrodynamical pre-SN evolution of RSGs and the simple expectation of hydrostatic models used to derive analytical scaling relations to infer the properties of supernovae, such as derived by \citep{popov1993a,kasen_2009_typeII,goldberg2020a}.

To this end, we use analytical homology relations and compute pre-SN stellar properties based on the hydrostatic $15\,\msun$ pre-SN model with radius $R_{\rm{HS}}$ and radial coordinate $r_{\rm{HS}}$. For each pulsation phase / radius $R_2$ shown in Fig.~\ref{fig:LC_vph_var_m}c, we compute a new envelope structure. From the base of the envelope to the surface, we compute new pre-SN structures by computing homology relations for the local radial coordinate $r_2$, density $\rho_2$, and temperature $T_2$ in every layer, applying an adiabatic ideal gas equation of state, which is a good first approximation for the convective envelope of a hydrostatic RSG \citep[e.g.,][]{kippenhahn_2013book}, such that:
\begin{eqnarray*}
r_2 = r_{\rm{HS}}\frac{R_2}{R_{\rm{HS}}}, &
\rho_2 =\rho_{\rm{HS}}\left(\frac{R_2}{R_{\rm{HS}}}\right)^{-3}, &
\rm{and}\,T_2 = T_{\rm{HS}}\left(\frac{R_2}{R_{\rm{HS}}}\right)^{-2}.
\end{eqnarray*}
Based on these homologously expanded/contracted models, we then compute the SN LC with \texttt{SNEC}. As the core properties are unchanged, the explosion properties are also assumed to remain the same, and we apply the same explosion parameters obtained from the \citet{muller2016a} model as for Fig.~\ref{fig:LC_vph_var_m}c (see Table~\ref{tab:summary-exp}).

As shown in Fig~\ref{fig:comp_SNLC_homology}, the re-structuring of the envelope in the hydrodynamic models (full lines) indeed leads to a significantly different SN LC shape and photospheric velocity evolution compared to the homologously rescaled models (dotted lines). The same trend of increased SN luminosity  for more extended models is recovered, but the scaled hydrostatic models all conserve their distinct type II-P like shape, and the more extended models have a longer plateau duration. In contrast, the SN LCs of the hydrodynamic models show significantly different shapes, ranging from type II-L-like to type II-P-like shapes. Interestingly, the time of the transition to the radioactive tail appears to be very similar for all hydrodynamical models, at about ~75d. Besides, no features are observed in the SN LCs of the re-scaled hydrostatic models in Fig~\ref{fig:comp_SNLC_homology}, as the SN progenitors do not exhibit the density inversions found in the hydrodynamic model caused by layers moving in- and outward due to the $\kappa\gamma$-mechanism (see also \citetalias{bronner2025} for a detailed discussion of this phenomenon). Besides, the ``early excess'' identified in Fig~\ref{fig:LC_vph_var_m} is not visible in the rescaled hydrostatic models. However, for the smallest radii, the SN LC of the hydrodynamic and rescaled hydrostatic models are very similar and resemble a classical type II-P like shape, except at very early times (less than 5d). For these models, the plateau duration and luminosity are all very similar.

Similarly, the photospheric velocity evolution is also systematically different for the scaled hydrostatic pre-SN structures compared to the the hydrodynamic pre-SN models, as shown in the lower panel of Fig~\ref{fig:comp_SNLC_homology}, especially for the more extended progenitors, which differ by a factor of up to two in the first 10 days. The photospheric velocities of the scaled HS model are all very similar and closely follow the trend of the initial HS model, as expected. Observing the early photospheric velocity evolution therefore provides a particularly strong constraint on the pre-SN structure of RSGs that experience a restructuring of their envelopes due to envelope pulsations.

As analytical scaling relations of SN LCs are all based on the assumption of a hydrostatic progenitor star \citep{popov1993a,kasen_2009_typeII,goldberg2019a}, applying these to SNe that experience a restructuring of their envelope due to pulsations is therefore inadvisable. For example, if the most extended progenitor in Fig.~\ref{fig:comp_SNLC_homology} were interpreted as a short-plateau II-P, one would find up to about 60\% difference in plateau duration.
In summary, this comparison emphasizes the importance of hydrodynamic pre-SN models to capture the full diversity of SN LCs and photospheric velocity evolutions from exploding RSGs.

\bibliographystyle{aasjournal}
\bibliography{references}

\end{document}